\documentstyle[aas2pp4]{article}

\begin{document}

\title{EUV SUNSPOT PLUMES OBSERVED WITH SOHO}

\author{P. Maltby, N. Brynildsen, P. Brekke, S. V. H. Haugan, O. Kjeldseth-Moe, \O. Wikst\o l} 
\affil{Institute of Theoretical Astrophysics, University of Oslo, 
P.O. Box 1029, Blindern, 0315 Oslo, Norway}

\and

\author{T. Rimmele}
\affil{National Solar Observatory, 
Sacramento Peak, Sunspot, NM 88349, USA}

\begin{abstract}

Bright EUV sunspot plumes have been observed in five out of 
nine sunspot regions with the Coronal Diagnostic Spectrometer 
-- CDS on SOHO. 
In the other four regions the brightest line emissions
may appear inside the sunspot but are mainly concentrated in small 
regions outside the sunspot areas.
These results are in contrast to those obtained during the
{\it Solar Maximum Mission}, but are compatible with the {\it Skylab} 
mission results.
The present observations show that sunspot plumes are formed in the upper 
part of the transition region, occur both in magnetic unipolar-- and 
bipolar regions, and may extend from the umbra into the penumbra.

\end{abstract}


\section{Introduction}

Based on monochromatic EUV images obtained during the {\it Skylab} 
mission Foukal et al. (1974) introduced the notation ``sunspot plumes'', 
defined as areas above sunspot umbrae that are ``the brightest features 
in an active region by an order of magnitude''.  
This led to the idea that sunspot plumes are regions within large magnetic 
loops, extending to altitudes of several thousand kilometers above the 
photosphere, in which the temperature is one to two orders of magnitude 
lower than in the corona of the surrounding active region 
(Noyes et al. 1985).
In contrast, Brueckner and Bartoe (1974) observed enhanced line emission 
both over plages and sunspots and Cheng, Doschek, \& Feldman (1976)
found that emission lines formed at temperatures below 
2.4 $\times$ 10$^5$K showed the same line emission 
over the sunspot as over the quiet network.
Evidence against the importance of sunspot plumes came from 
the UVSP instrument on the {\it Solar Maximum Mission - SMM}. 
Kingston et al. (1982) found that the emission measures over the 
penumbrae were higher than over the umbrae.
The {\it SMM} observations led Gurman (1993) to conclude that the umbral 
transition region was generally indistinguishable from the quiet 
transition region.

There appears to be at least two possible explanations for the discrepancy 
between the observations obtained with the S055 {\it Skylab} and the 
UVSP {\it SMM} instruments.
One possibility is that sunspot plumes are formed in the upper part of 
the transition region and therefore are easier observed 
with S055 than with UVSP.
Another possibility is that one type of sunspots have plumes whereas
others do not.
The present observations show that sunspot plumes occur both in magnetic 
unipolar and bipolar active regions and are most apparent in emission 
lines formed in the upper part of the transition region.

\begin{table}[htb]
\caption{Observed active regions}
\begin{center}%\scriptsize
\begin{tabular}{llccr}
\tableline
NOAA & Date    & $\theta$  & Mag.  \,\,\, \\
     &         & (degrees) & Class  \,\,\, \\
\hline
7973 & 1996 June 26          &   16    & A \,\,\, \\
7981 & 1996 August 2         &   16    & B \,\,\, \\   
7986 & 1996 August 29        &   17    & A \,\,\, \\    
7999 & 1996 November 28      &   36    & B \,\,\, \\   
8011 & 1997 January 16       &   13    & B \,\,\, \\     
8073 & 1997 August 16        &   15    & A  \,\,\, \\        
8076 & 1997 August 30        &   20    & B  \,\,\, \\        
8083 & 1997 September 9      &   35    & B  \,\,\, \\        
8085 & 1997 September 15     &   44    & B  \,\,\, \\        
\tableline
\end{tabular}
\end{center}
\tablenotetext{}{The angle $\theta$ = heliocentric angle. Magnetic field classification:}
\tablenotetext{}{A = unipolar, B = bipolar.}
\end{table}

\section{Observations and Data Reduction}

Observations of nine sunspot regions (see Table 1) were obtained with
the Normal Incidence Spectrometer (NIS) of the Coronal Diagnostic 
Spectrometer -- CDS (Harrison et al. 1995), as part of a joint
observing program on the Solar and Heliospheric Observatory -- SOHO. 
A large fraction of the observing time was used to raster an area of 
120$\arcsec \times$ 120$\arcsec$, moving the narrow 2.0$\arcsec$ 
spectrometer slit perpendicular to the slit direction in steps of 2.0$\arcsec$.
The exposure time was 20 s, each raster was recorded during
25 min and contains information from 60 adjacent locations for ten emission 
lines. 
Table 2 gives the line list and the corresponding ionization temperatures. 
For each observing sequence several rasters, up to thirteen, were 
obtained.

\begin{table}[htb]
\caption{Selected spectral lines} 
\begin{center}%\scriptsize
\begin{tabular}{lrrll}
ID & $\lambda$ ({\AA}) & log T (K) &\\
\tableline
He I    & 522.2 & 4.3 & & \\
He I    & 584.3 & 4.3 & & \\
O III   & 599.5 & 5.0 & & \\
O IV    & 554.5 & 5.2 & & \\
O V     & 629.7 & 5.4 & & \\
Ne VI   & 562.8 & 5.6 & & \\
Mg VIII & 315.0 & 5.9 & & \\
Mg IX   & 368.0 & 6.0 & & \\
Fe XIV  & 334.1 & 6.3 & & \\
Fe XVI  & 360.7 & 6.4 & & \\
\tableline
\end{tabular}
\end{center}
\end{table}

The data acquisition and detector characteristics that are relevant
for this study were described by Harrison et al. (1995).
Briefly, the CDS data are corrected for geometrical distortions, 
the CCD readout bias is removed, the non-wavelength-dependent 
calibration parameters peculiar to the detector are applied, including 
the exposure time, the amplification of the microchannel plate, and 
a flat-field correction.  
The final step in calibrating is to convert the photon events into 
absolute units.
The line parameters, peak intensity, wavelength shift and line width
are determined by a least squares fit to the observations, see
Brynildsen et al. (1997). 
The data material consists of line profiles that are well represented 
by a single Gaussian shape. 
Small regions with complicated line profiles and regions with rapid 
time evolution are outside the scope of this paper.

The CDS images were coaligned with white light images using
magnetograms observed with the Michelson Doppler Imager 
(Scherrer et al. 1995) from SOHO.
To determine the location of regions with peak line intensity 
$I > I_{p}$, where $I_{p}$ is a preselected value of $I$, we
introduce the notations:

\noindent
$F_{U} (I > I_{p})$ = fraction of umbra covered with 
$I > I_{p}$, 

\noindent
$F_{S} (I > I_{p})$ = fraction of sunspot covered with 
$I > I_{p}$, 

\noindent
$f_{U} (I > I_{p})$ = fraction of area with $I > I_{p}$ 
located above the umbra,

\noindent
$f_{S} (I > I_{p})$ = fraction of area with $I > I_{p}$ 
located above the sunspot.

\noindent
The brightest features, such as the sunspot plumes, will be located by
$I > I_{p} = 5\overline{I}$, where $\overline{I}$ is the average 
peak line intensity value within the rastered area, 
120$\arcsec \times$ 120$\arcsec$.

\section{Results}

In Figures 1 and 2 (Pl.00 and 00) the brightest features with
peak line intensity $I > 5\overline{I}$ are encircled by 
yellow contours, whereas medium bright features with 
$I > 2.5\overline{I}$ are encircled by green contours.
Below we give our reasons for identifying the brightest feature in 
NOAA 7986, observed on 29 August 1996 (see Fig.~1), with 
a sunspot plume.
In contrast, Figure~2 shows that nearly all the brightest features in 
NOAA 7999, observed on 28 November 1996, are located outside the sunspot.
The size and the location of the brightest emission features with
peak line intensity $I > 5\overline{I}$ are given in Table 3
for the entire set of observations.
We limit the list to the three emission lines where the
sunspot plumes are most apparent.

An excellent illustration of a sunspot plume is presented in Figure~2 
of Foukal et al. (1974), based on observations of McMath region 12543.
To compare the brightest emission feature in NOAA 7986 (Fig.~1)
with that of a sunspot plume let us first consider the variation
with the line formation temperature. 
The line emission in the sunspot plume in McMath region 12543
exceeds that above the adjacent plage region 
in O~{\scriptsize IV} 554~{\AA}, O~{\scriptsize VI} 1032~{\AA}, and 
Ne~{\scriptsize VII} 465~{\AA}, see Figure~3 in Foukal et al. (1974). 
Figure~1 shows that the brightest emission above the umbra in NOAA 7986 
is observed in a compatible temperature range since the brightest
emission is observed in O~{\scriptsize IV} 554~{\AA}, 
O~{\scriptsize V} 629~{\AA}, and Ne~{\scriptsize VI} 562~{\AA}.
Next consider the spatial extent of the brightest emission.
According to Foukal et al. (1974) the plume's  
``minimum half-width is definitely smaller than the 
diameter of the umbra''.
However, their Figure~4 shows that the spatial extent changes 
from one emission line to another.
In Ne~{\scriptsize VII} 465~{\AA}, where the plume is most 
apparent, the FWHM of the plume exceeds 30$\arcsec$ which is 
considerably larger than the umbral diameter ($\approx$ 10$\arcsec$). 
We find that the size of the bright features in the sunspots of 
NOAA 7986 and McMath region 12543 are compatible.
Based on these comparisons we conclude that the sunspot in 
NOAA 7986 (Fig.~1) shows a sunspot plume.

A sunspot plume is located above the sunspot and is the brightest feature 
within the active region.
It is important to note that both the extent and the location of enhanced 
line emission depend on the preselected intensity level $I_{p}$. 
We show in Figure~3 the fraction, $f_{S} (I > I_{p})$, of the area 
with peak line intensity $I > I_{p}$ which is located above the sunspot 
as function of the intensity ratio, $I_{p}/ \overline I$.
For several active regions we find that as $I_{p}/ \overline I$ increases
towards the brightest features an increasing fraction of the line emission 
in O~{\scriptsize IV} 554~{\AA}, O~{\scriptsize V} 629~{\AA}, and 
Ne~{\scriptsize VI} 562~{\AA} is located inside the sunspot.
This means that in a search for sunspot plumes one cannot simply look for
enhanced line emission, but must select the brightest regions by
choosing a criterion, such as, $I > I_{p} = 5\overline{I}$.

If it is required that the sunspot plume must be positioned directly 
above the umbra, only two of the nine sunspots contain a sunspot plume.
Taking into account the measured size of sunspot plumes a more reasonable
requirement is that a plume is located in an area above the sunspot 
that includes the umbra or parts thereof and may extend into the penumbra.
With this requirement the following five sunspots show plumes in 
O~{\scriptsize V} 629~{\AA} and Ne~{\scriptsize VI} 562~{\AA}:
NOAA 7973, 7986, 8011, 8073, and 8085.
In these sunspots the plumes are centered in the 
penumbra, the umbra (see Fig.~1), the umbra, the rim of the umbra
and the penumbra, respectively.
It is possible that also NOAA 8076 should be regarded as containing
sunspot plumes since two of the brightest line emission regions in 
O~{\scriptsize V} 629~{\AA} and Ne~{\scriptsize VI} 562~{\AA} are located
within the largest, leading sunspot and the third bright line emission 
region covers most of a following sunspot, but extends outside the
sunspot. 
The other three active regions show bright emission both inside
and outside the sunspot in Ne~{\scriptsize VI} 562~{\AA}.

\begin{table}[htbp]
\caption{Size and location of the brightest emission regions in O~{\scriptsize IV} 554~{\AA}, O~{\scriptsize V} 629~{\AA} and Ne~{\scriptsize VI} 562~{\AA}}
\begin{center}\scriptsize
\begin{tabular}{llrrrrrrr}
\tableline
NOAA & ID &size &$f_{U}$&$f_{S}$&$F_{U}$&$F_{S}$& \\
     &    &(arc sec)$^2$& & $I > 5\overline{I}$  &       &       & \\
\tableline
7973 & O IV   &   7 & 0.00 & 0.00 & 0.00 & 0.00 & \\
     & O V    &  31 & 0.11 & 0.67 & 0.14 & 0.04 & \\
     & Ne VI  &  99 & 0.03 & 0.48 & 0.14 & 0.08 & \\
\tableline
7981 & O IV   & 384 & 0.03 & 0.42 & 0.03 & 0.10 & \\
     & O V    & 493 & 0.03 & 0.63 & 0.04 & 0.19 & \\
     & Ne VI  & 248 & 0.00 & 0.59 & 0.00 & 0.09 & \\
\tableline
7986 & O IV   & 105 & 0.26 & 1.00 & 0.80 & 0.21 & \\
     & O V    & 218 & 0.16 & 0.91 & 1.00 & 0.39 & \\
     & Ne VI  & 265 & 0.13 & 0.86 & 1.00 & 0.46 & \\
\tableline
7999 & O IV   & 156 & 0.00 & 0.00 & 0.00 & 0.00 & \\
     & O V    & 129 & 0.00 & 0.00 & 0.00 & 0.00 & \\
     & Ne VI  & 122 & 0.25 & 0.25 & 0.05 & 0.02 & \\
\tableline
8011 & O IV   &   0 & 0.00 & 0.00 & 0.00 & 0.00 & \\
     & O V    &  14 & 0.75 & 0.75 & 0.04 & 0.04 & \\
     & Ne VI  &  24 & 0.29 & 0.29 & 0.03 & 0.03 & \\
\tableline
8073 & O IV   &   3 & 0.00 & 1.00 & 0.00 & 0.01 & \\
     & O V    &  37 & 0.36 & 0.91 & 0.17 & 0.07 & \\ 
     & Ne VI  &  78 & 0.22 & 0.61 & 0.21 & 0.10 & \\
\tableline
8076 & O IV   &  54 & 0.06 & 0.12 & 0.03 & 0.01 & \\
     & O V    & 221 & 0.20 & 0.62 & 0.45 & 0.14 & \\ 
     & Ne VI  & 238 & 0.16 & 0.29 & 0.38 & 0.07 & \\
\tableline
8083 & O IV   & 377 & 0.06 & 0.26 & 0.08 & 0.08 & \\
     & O V    & 262 & 0.08 & 0.26 & 0.07 & 0.06 & \\ 
     & Ne VI  & 102 & 0.07 & 0.40 & 0.02 & 0.03 & \\
\tableline
8085 & O IV   & 102 & 0.10 & 1.00 & 0.07 & 0.09 & \\
     & O V    & 282 & 0.23 & 0.93 & 0.46 & 0.22 & \\ 
     & Ne VI  & 333 & 0.12 & 0.76 & 0.29 & 0.21 & \\
\tableline
\end{tabular}
\end{center}
\end{table}

In contrast to the {\it SMM} observations of Gurman (1993), the
present observations confirm the existence of sunspot plumes.
The {\it SMM} observations were obtained closer to a sunspot maximum 
than the present observations.
We cannot exclude a selection effect since the sunspots listed by 
Gurman (1993) are all magnetic bipolar, whereas unipolar sunspots 
were observed both by Foukal et al. (1974) and by us (see Table~1).
However, this cannot be the entire explanation since we also 
observe sunspot plumes in bipolar sunspots.
It appears that the CDS instrument is well suited to measure 
emission lines formed in the upper part of the transition region,
where the sunspot plumes are most apparent.
The higher sensitivity of the CDS instrument than the UVSP {\it SMM} 
instrument to line emission in this region of the solar atmosphere 
may be one of the reasons for the difference in results between the two 
instruments.
The present observations suggest that sunspot plumes are formed in 
the upper part of the transition region, occur both in magnetic 
unipolar and bipolar regions, and may extend outside the umbra and 
into the penumbra. 

\acknowledgements 

We would like to thank all the members of the large international CDS 
team for their extreme dedication in developing and operating this 
excellent instrument, the Michelson Doppler Imager team for permission 
to use their data for coalignment purposes and the Research Council of
Norway for financial support. Data from Mees Solar Observatory, 
University of Hawaii, are produced with the support of NASA grant 
NAG 5-4941 and NASA contract NAS8-40801.
SOHO is a mission of international cooperation between ESA and NASA.

%\clearpage

\begin{figure*}[p] %PLATE
 \figcaption[]{ Images of peak line intensities in NOAA 7986 observed 
on August 29, 1996. 
Regions with enhanced intensity are shown as dark orange regions. 
Areas with peak line intensity, $I$, larger than 2.5 and 5 times the average 
intensity, $\overline I$, are encircled by green and yellow contours.
The images are ordered after increasing line formation temperature, 
starting in the upper left hand corner. The contours of the umbra and
penumbra are from white light observations at the Mees Solar Observatory, 
Haleakala, Hawaii. The scales in arc sec are in a reference system
where the origin coincides with the centre of the solar disk.}
\label{fig1}
\end{figure*}

\begin{figure*}[p] %PLATE
 \figcaption[]{Images of peak line intensities in EUV
emission lines observed on 28 November 1996.
The same color code etc. as in Figure~1.  
White light image of NOAA 7999 observed the previous day 
with the Vacuum Tower Telescope at the National
Solar Observatory, U.S.A. is shown (top). 
Only minor changes in the sunspot contour were observed from one day 
to the next.}
\label{fig2}
\end{figure*}

\begin{figure*}[htb]
\figcaption[]{ The fraction, $f_{S} (I > I_{p})$, of the area 
with peak line intensity $I > I_{p}$ that is located above the sunspot 
as a function of the intensity ratio $I_{p}/ \overline{I}$.
Note the tendency for some sunspots to show the brightest emission
above sunspots in O~{\scriptsize IV} 554~{\AA}, 
O~{\scriptsize V} 629~{\AA}, and Ne~{\scriptsize VI} 562~{\AA}.}
\label{fig3}
\end{figure*}

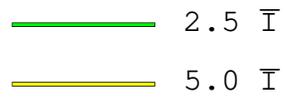
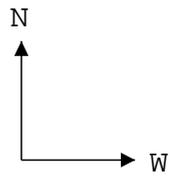
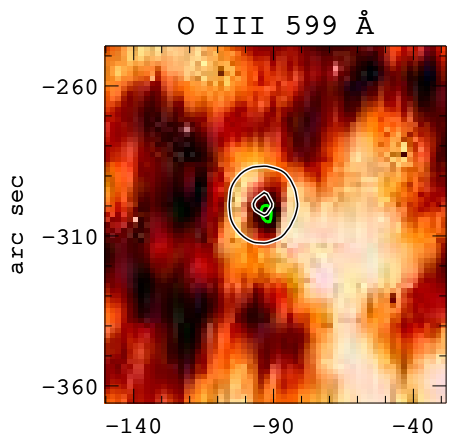
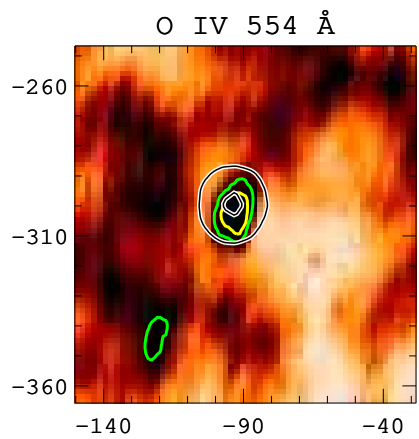
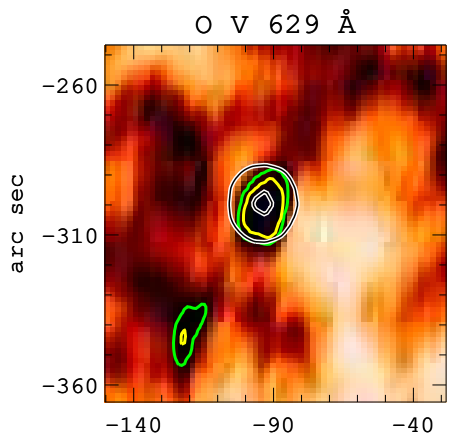
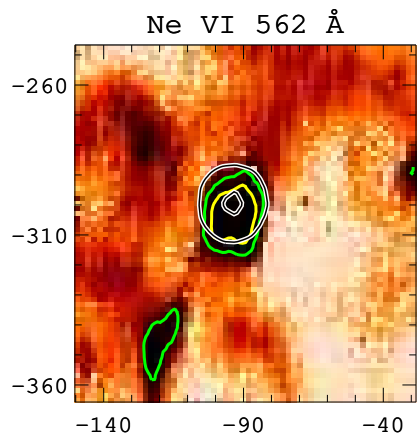
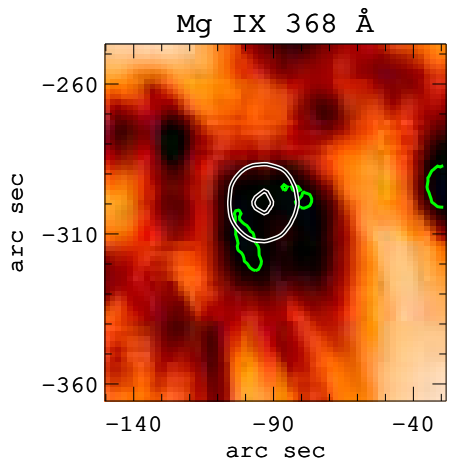
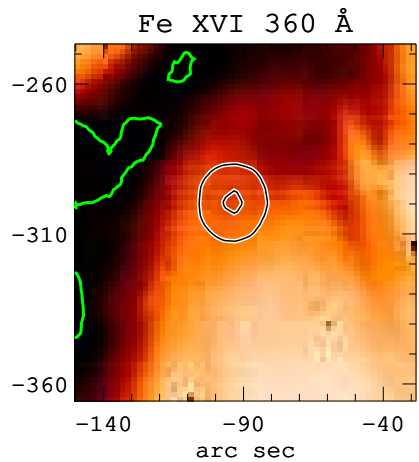
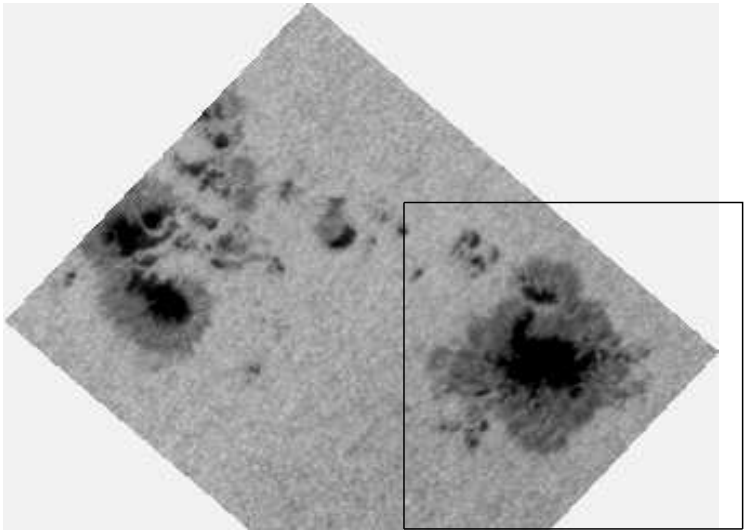
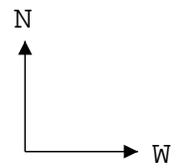
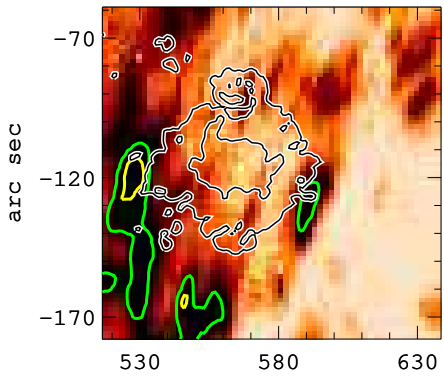
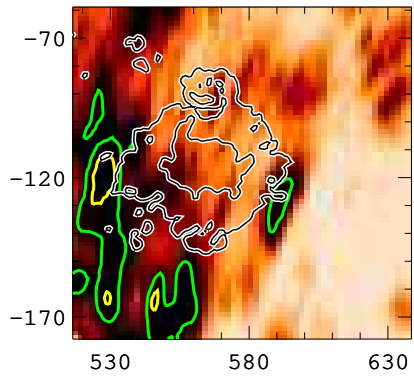
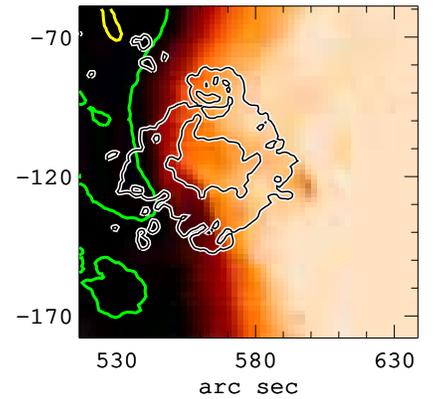
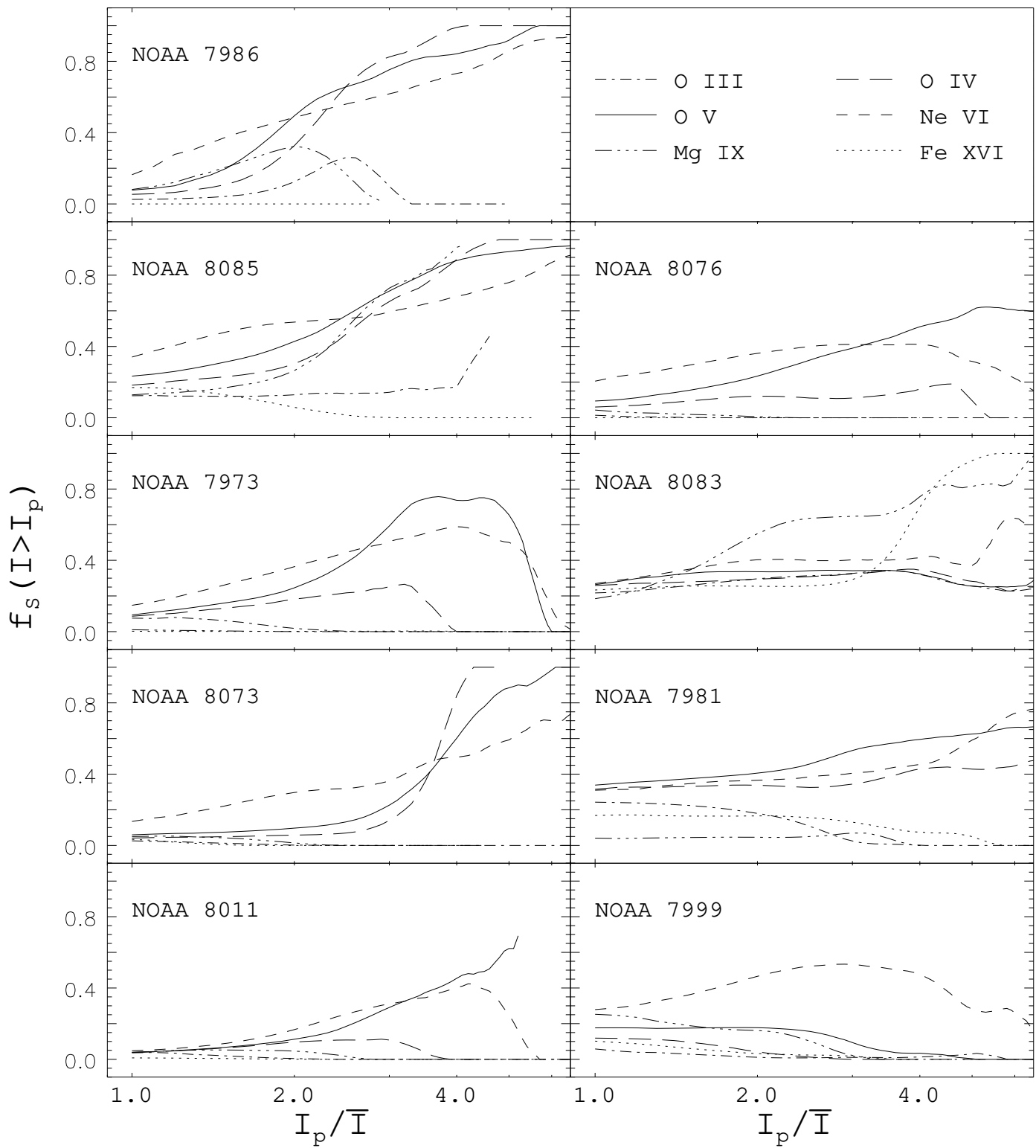
\end{document}